\documentclass[11pt,singlespacing]{article}

\usepackage{amsfonts}
\usepackage{graphicx}
\usepackage{graphics}
\newcommand{\R}{\mathbb{R}}
\newcommand{\Z}{\mathbb{Z}}
\newcommand{\Q}{\mathbb{Q}}

\newcommand{\U}{\mathbb{U}}

\begin{document}

\title{The Haar Wavelet Transform of a Dendrogram: Additional Notes}
\author{Fionn Murtagh\thanks{F. Murtagh is with the Department 
of Computer Science,
Royal Holloway, University of London, Egham, Surrey, TW20 0EX, England.  
Email fmurtagh@acm.org}
}
\markboth{Haar Wavelet Transform of a Dendrogram: Additional Notes}{Haar Wavelet 
Transform of a Dendrogram: Additional Notes}
\maketitle

\begin{abstract}
We consider the wavelet transform of a finite, 
rooted, node-ranked, $p$-way tree, focusing on the case of binary ($p = 2$)
trees.  We study a Haar wavelet transform on this tree. 
Wavelet transforms allow for multiresolution analysis
through translation and dilation of a wavelet function.  We explore how this
works in our tree context. 
\end{abstract}


\section{Introduction}

In a companion paper, which we will refer to as Paper I (``The Haar wavelet
transform of a dendrogram''),   
a new transform is applied to a 
hierarchical clustering.  Various examples are given of uses of this
transform, prior to applying the inverse transform.  

In this paper, we look at linkages with other ways of understanding 
the wavelet transform, with the classical d\'emarche described in 
Appendix 1.  Our aim is to understand the wavelet transform when 
applied to hierarchical clustering dendrograms (where notation and 
expression as ultrametric topology are summarized in Appendix 2).

After all, both the wavelet transform and hiearchical clustering aspire to 
multiresolution or multiscale analysis.  The natural question is then: 
how do they differ and are there different aspects that they bring to the 
data analysis task? 

A good deal of recent work on wavelet transform has been through 
group theoretic approaches.  

Foote et al.\ (2000a) point to how
group theoretic understanding can lead to a ``wealth of new analysis filters'' 
(in the context of multiresolution signal and image analysis).  The 
same point is made by the SMART project (SMART, 2005), including the 
change to have automatic generation of new transform algorithms.  Believing
that algorithms should be developed if and only if the there is a 
verifiable user need for them, we would instead point to another reason 
why group theoretic understanding is crucial to data analysis.  A great deal 
of observed reality can be understood by way of observed symmetries, and 
groups summarized and encapsulate the properties of these symmetries.  For
time evolving phenomena, therefore, or spatial coordinate referenced 
phenomena, it may be possible to replace analysis that is time-referenced
or referenced to particular coordinate systems with a more general, more
generic, symmetry analysis.  This is the vision opened up by the study of 
group actions on a set of objects.   

Parenthically, one fascinating way as to how this works can be seen in 
Cendra and Marsden (2003).  The authors develop (i) an analytic theory 
of dynamics 
as functions of spatial and temporal coordinates; and (ii) group theoretic
interpretation of, in parts of the study, return or phase maps.

Our approach can be stated as follows.  Let $\U$ be an 
ultrametric space, associated with an $m$-dimensional embedding, $\R^m$.  
We note that an ultrametric space is necessarily of 0 dimenensionality; 
and that minimal dimensionality real embedding of an ultrametric has been 
studied by Lemin and others (Lemin, 2001; Bartal et al., 2004).

The partial order of (clopen) set inclusion is denoted
by the binary tree or hierarchy, $H$.  Consider the group action 
comprising rotations or cyclic permutations (these are equivalent) of 
subnodes of any node in $H$, and we will denote this group as $G_H$.  Then 
we study the wavelet transform of $L^2(\U)$ resulting from the actions of 
group $G_H$.  

Having already discussed the new wavelet transform in Paper I, we can 
give one result relating to it in the context of the group of equivalent 
representations of $H$ as follows.

{\em Theorem:}  For all $2^{n-1}$ equivalent representations of $H$
(here: unlabeled graph isomorphisms), the 
dendrogram Haar wavelet transform is unique.

The proof follows from the definition of the wavelet coefficents at each
level, $\nu$; whereas the equivalent representations of $H$ are intra level.

It follows from this theorem that we have a unique matrix representation 
of a dendrogram.  

\section{Previous Work on Wavelet Transforms of Data Tables}
\label{sect3}

In this section we will review recent work using wavelet transforms on
data tables, and show how our work represents a radically new approach to
tackling similar objectives.

Approximate query processing arises when data must be kept
confidential so that only aggregate or macro-level data can be divulged.
Approximate query processing also provides a solution to access of
information from massive data tables.

One approach to approximate database querying through aggregates is
sampling.  However a join operation applied to two uniform
random samples results in a non-uniform result, which furthermore is
sparse (Chakrabarti, Garofalakis, Rastogi and Shim, 2001).
A second approach is to keep histograms on the coordinates.  For a
multidimensional feature space, one is faced with a ``curse of
dimensionality'' as the dimensionality grows.  A third approach is
wavelet-based, and is of interest to us in this article.

A form of progressive access to  the data is sought, such that aggregated
data can be obtained first, followed by greater refinement of the data.
The Haar wavelet transform is a favored transform for such purposes,
given that reconstructed data at a given resolution level is simply
a recursively defined mean of data values.  Vitter and Wang (1999)
consider the combinatorial aspects of data access using a Haar wavelet
transform, and based on a multi-way data hypercube.  Such data, containing
scores or frequencies, is often found in the commercial data mining context
of OLAP, On-Line Analytical Processing.

As pointed out in Chakrabarti et al.\ (2001), one can treat multidimensional
feature hypercubes as a type of high dimensional image, taking the
given order of feature dimensions as fixed.  As an alternative a
uniform ``shift and distribute'' randomization can be used
(Chakrabarti et al., 2001).

There are problems, however, in directly applying a wavelet transform
to a data table.  Essentially, a relational table (to use database
terminology; or matrix) is treated in the same way as a 2-dimensional
pixelated image, although the former case is invariant under row and column
permutation, whereas the latter case is not (Murtagh, Starck and Berry, 2000).
Therefore there are immediate problems related to
non-uniqueness, and data order dependence.

What if, however, one organizes the data such that adjacency has
a meaning?  This implies that similarly-valued objects, and/or
similarly-valued features, are close together.  This is what we do,
using any hierarchical clustering algorithm (e.g., the Ward or minimum
variance one).  

Without loss of generality, as seen in these figures, we assume that a
hierarchy is a binary, rooted tree; and equivalently that the series of
agglomerations involve precisely two clusters (possibly singleton clusters)
at each of the $n - 1$ agglomerations where there are $n$ observations.
These $n$ observations are usually represented by $n$ row vectors in our
data table.

A significant advantage in regard to hierarchical clustering is that
partitions of the data can be read off at a succession of levels, and this
obviates the need for fixing the number of clusters in advance.  All
possible clustering outcomes are considered.  (Remark: of course, relative
to any one of the commonly used cluster homogeneity criteria, each
partition is guaranteed to be sub-optimal at best.)

\section{The Haar Wavelet Transform of a Dendrogram: Summary}
\label{sect2}

In this article, we will denote the agglomeration of two clusters, $q$ and 
$q'$, as cluster $q''$.  So (left or right subtree) nodes in the 
dendrogram are associated with the child (elder or younger) subnodes.  We 
can define the elder cluster as $q$ such that $\nu(q) > \nu(q')$, but we will
{\em not} be concerned with whether or not elder corresponds to left, and 
younger to right.  

For $n$ objects or observation vectors, another notation that we can use
is that the hierarchy $H$ is the set of clusters indexed from 1 to $n$: 
$H = \{ q_1, q_2, \dots , q_{n-1} \}$.  We will always assume in this 
article, for convenience of exposition and with little loss of generality, 
that for distinct clusters $\nu(q) \neq \nu(q')$.  

Whenever the distinction between the following become important, we will
clearly distinguish between them: clusters; nodes; sets of objects; 
sets of indices of objects; and p-adic number representation of indices
of objects.   

The Haar algorithm, as discussed in Paper I, is as follows:

\begin{enumerate}

\item Take each cluster $q''$ in turn, proceeding in sequence through 
$q'' \in \{ q_1, q_2, \dots , q_{n-1} \} $.  
\item Apply the smoothing function, $s$: $s(q'') = \frac{1}{2}(q + q')$.
\item Thereby apply the detail function, $d$: $d(q'') = 
s(q'') - s(q') = - (s(q'') - s(q)) $.  
\item Return to step 1 until all $n-1$ clusters are processed.
\end{enumerate}

For details of how the clusters also take terminal nodes (objects) into 
account, see Paper I.   

Now, it is clear from construction that perfect reconstruction of the 
input data (alternatively expressed, perfect undoing of the foregoing Haar 
algorithm) is guaranteed, given all of the following: 
(i) all of the detail function 
values, (ii) the final smooth, $s(q_{n-1})$, (iii) the definition of 
the dendrogram, and (iii) a convention of 
left and right subtree that allows us to traverse down the tree from $q''$ to 
both $q$ and $q'$.

In practice our objectives are to explore the foundations of two distinct
approaches.  Both seek a Haar wavelet basis.  These two approaches are
as follows and can express the 2 input data cases considered in section 
4.3 (``The Input Data'')  of Paper I.  

\begin{itemize}
\item Wavelet transform in an ultrametic topology: 
Induce the Haar basis from 
the hierarchy $H$ that expresses the relationships in 
a set of ultrametrically related points, $I$.

\item Wavelet transform on embedded subsets: 
Induce the Haar basis from 
the hierarchy $H$ defining a set of subsets of $I$.
\end{itemize}

In the ultrametric case, each point $i \in I$ defines an
$m$-dimensional vector: $i \in \R^m$.  For notational 
convenience therefore $i$ is 
either the index, or a vector.

In the set of subsets case, each point $i \in I$ can be defined as an 
$n$-dimensional index vector.  Thus for example 
the sequentially second point is defined as 
$(0, 1, 0, 0, \dots , 0)$.  

Both practical cases above can be expressed as follows: we carry out a 
wavelet transform in $L^2(G)$ where $G$ is the group of alternative 
representations of a given hierarchy, $H$.  The points $i \in I$ are 
associated either with vectors in $\R^m$ or with an orthonormal vector 
set in $\Z^n$.  (Note how $\Z^n$ is $n$-dimensional, whereas $\R^m$ is 
$m$-dimensional.  The cardinality of $I$ is $n$.  The dimensionality of 
a feature or attribute space is $m$.)

\section{Wavelet Transform on Discrete Fields}
\label{discrete}

In this section we look at the wavelet transform on discrete fields, and 
in particular on $\ell^2(\Z_p)$.  This is realized through cyclic 
representations of the affine group of $\Z$ or of $\Z_p$.  

In wavelets, we are seeking a representation of our data which has 
``covariant'' properties relative to scale: for example, for tranlation
``covariance'', the representation of a shifted signal must be a shifted
copy of the representation of the signal (Torr\'esani, 1994).  In the 
group theory perspective, such ``covariance'' properties (i.e.\ with 
respect to the action of a symmetry group) are the starting 
point, and the representation is to be derived from them.  The ``covariance''
group ``turns out to be isomorphic (up to a compact factor) to the geometric
phase space of the representation'' (Torr\'esani, 1994, p.\ 6).

Traditionally, the wavelet transform is covariant with respect to a 
group action applicable to images, signals, time series, etc., viz.\ the 
affine group of the real line, which is a continuous group (Antoine et al.,
2000a).  Thus a first task is to bypass the need for a continuous group.

The group law of the affine group, in generic form $ax + b$, generates 
translations and dilations.  The action of the $ax + b$ group on $\R$
means: $(a,b) : x \longrightarrow ax + b$.  We have the following product: 

\begin{equation}
 (b, a) \cdot (b', a') = (b + a b', a a')
\label{eqntd}
\end{equation}

Here, the identity is: $(1, 0)$.  The inverse of $(a,b)$ is: 
$(a, b)^{-1} = (a^{-1}, - \frac{b}{a})$.  This is a non-commutative Lie 
(and thus continuous) group.

Flornes et al.\ (1994) 
consider a discrete wavelet transform to begin with,
specified on the Hilbert space $\ell^2(\Z_p)$ (where $\Z$ are the integers,
$\Z_p$ are integers mod $p$ where $p$ is prime 
for reasons explained below; and 
$\ell^2$ implies finite energy from discrete values, or being square
integrable).  The Haar measure is defined on locally compact groups, 
permitting integration over group actions or members; and a locally 
compact separable group is consistent with the square integrable property.
The group at issue here is the cyclic representation of the affine group; 
or its finite analog, the affine group mod p (Foote et al., 2000a).

A discussion of square integrable group representations in the 
context of time-frequency transforms, including the continuous wavelet 
transform, can be found in Torr\'esani (2000); and Torr\'esani (1994) 
discusses the counterexample 
case of the rotation group, $S^2$, on the 2-dimensional sphere, 
which gives rise to a representation which is not square integrable.  
  
The wavelet transform considered by Flornes et al.\ (1994) has the 
following operations:

\begin{eqnarray}
\mathrm{Translation:} & T_b f(n) = f(n - b)  & 
       \mathrm{for} \ \ f \in \ell^2(\Z_p) \\  \nonumber
\mathrm{Dilation:} & D_a f(n) = f(a^{-1} n)  &  \\
\label{eqntd2}
\end{eqnarray}

The reason why $p$ has to be prime is as follows.  Consider the p-adic 
number representation of $\Z_p$.  For the p-adic respresentation of 
$\Z_p$ to be a field, i.e.\ to have an inverse, $p$ must be prime.  

The unitary representation of a group $G$ is a mapping into a (complex) 
Hilbert space.  
Flornes et al.\ 
(1994) define the following unitary representation, $\pi$, on the 
group, mapping into unitary operations on $\ell^2(\Z_p)$:

$$ \pi(g) f(n) = f(a^{-1} (n - b)) $$

In terms of the translation and dilation operators, we have:

$$\pi(b,a) = T_b D_a$$

Thus far, a purely discrete wavelet transform is at issue.  However if
we take our function values  $f$ defined on $\Z_p$ as sampled values from a 
continuous signal, then problems of interpolation arise.  It simply is not
good enough to transform our discrete data independently of awareness of 
the underlying continuum.  Note that this issue is at the nub of where 
data analysis differs from signal processing.  In data analysis, mostly 
a data cloud is taken as given (potentially leading to a 
combinatorial perspective) or as a stochastic realization (leading to a 
statistical modeling).  In signal processing, the observed data are 
samples of an underlying topological or other continuous structure. 
It is a prime objective in the signal processing context to keep the 
processing of the observed, sampled data fully and provably consistent with 
the underlying continuous structures.  

A way to address this issue of interpolation is 
to use B spline filters (the simplest example of which is 
the box function used by the Haar wavelet transform) to smooth the data, 
thereby ``filling the holes'' between gaps in the sampled values, before 
dilating.  This is termed pseudo-dilation.  

Relations (\ref{eqntd2}) can be re-expressed as follows (Torr\'esani, 1994):

\begin{eqnarray}
\mathrm{Translation:  } &
 T_b f(n) = f(n - b)   & \mathrm{for} \ \  f \in \ell^2(\Z_p) \\  \nonumber
\mathrm{Dilation:  } & D_a f(n) = f(a^{-1} n) & \mathrm{if} \ \ a  \ \
\mathrm{divides} \ \ n \\  \nonumber
         &  = 0            & \mathrm{otherwise}   \\
\label{eqntd3}
\end{eqnarray}

Then the affine multiplication law is verified by $\{ T_b, D_a \}$
(using relation \ref{eqntd}): 

\begin{equation}
T_b D_a T_{b'} D_{a'} = T_{b + ab'} D_{a a'} 
\end{equation}

When $f \in \ell^2(\Z_p)$ with $p$ prime, then $a$ always divides $n$. 
The affine group on $\ell^2(\Z_p)$ is well-defined; $T_b$ and $D_a$ define 
a representation of this group; and it turns out that the square integrable 
property holds.  

In general for $f \in \ell^2(\Z)$  the \`a trous algorithm is used 
incorporating both continuously defined dilation, and discretization of the 
function.  

We could embed each node of a hierarchical clustering, defined as we 
always do so as binary, rooted tree, in $\Z_2$.  Lang (1998) develops 
a wavelet transform approach (including the Haar wavelet transform and 
others) for such a 2-series local field, or Cantor
dyadic group.  Taking each cluster $q \in Q$ or node in the tree individually
is not satisfactory from our point of view, and so we look further for a 
more pleasing way to process a hierarchy.  

\section{Wavelet Bases on Local Fields}

Wavelet transform analysis is the determining of a ``useful'' 
basis for $L^2(\R^m)$ with the following properties:
\begin{itemize}
\item induced from a discrete subgroup of $\R^m$, 
\item using translations on this subgroup, and
\item dilations of the basis functions.
\end{itemize}

Classically (Frazier, 1999; Debnath and Mikusi\'nski, 1999; 
Strang and Nguyen, 1996) 
the wavelet transform avails of a wavelet function $\psi(x) \in
L^2(\R)$, where the latter is the space of all square integrable functions.
Wavelet transforms are bases on $L^2(\R^m)$, and the discrete lattice
subgroup
$\Z^m$ is used to allow discrete groups of dilated translation operators 
to be induced on $\R^m$.  Discrete lattice subgroups are typical of 
2D images (the lattice is a pixelated grid) or 3D images (the lattice is 
the voxelated grid) or spectra or time series (the lattice is the set of 
time steps, or wavelength steps).  

Sometimes it is appropriate to consider the construction of wavelet 
bases on $L^2(G)$ where $G$ is some group other than $\R$.  In 
Foote, Mirchandani, Rockmore, Healy and Olson (2000a, 2000b; see also 
Foote, 2005) this is done
for the group defined by a  quadtree, in turn derived from a 2D image. 
To consider the wavelet transform approach not in a Hilbert space
but rather in locally-defined and discrete spaces we have to change the 
specification of a wavelet function in $L^2(\R)$ and instead use
$L^2(G)$.   

Benedetto (2004) and Benedetto and Benedetto (2004) considered in detail  
the group $G$ as a locally compact abelian group.  
Analogous to the integer grid, $\Z^m$, a compact subgroup is used to allow a 
discrete group of operators to be defined on $L^2(G)$.  
The property of locally compact (essentially: finite and free of edges)
abelian (viz., commutative) groups that is most important is the 
existence of the Haar measure (Ward, 1994).  The Haar measure allows
integration, and definition of a topology on the algebraic structure of 
the group.

Benedetto (2004) considers the following cases, among others, of wavelet bases 
constructed via a sub-structure: 

\begin{itemize}
\item Wavelet basis on $L^2(\R^m)$ using translation operators defined
on the discrete lattice, $\Z^m$.  This is the situation discussed above,
which holds for image processing, signal processing, most time series 
analysis (i.e., with equal length time steps), spectral signal processing,
and so on.  As pointed out by Foote (2005), this framework allows the 
multiresolution analysis in $L^2(\R^m)$ to be generalized to $L^p(\R^m)$
for Minkowski metric $L^p$ other than Euclidean $L^2$. 
\item Wavelet basis on $L^2(\Q_p)$, where $\Q_p$ is the p-adic field, 
using a discrete set of translation operators.  This case has been
studied by Kozyref, 2002, 2004; Altaisky, 2004, 2005.  See also the
interesting overview of Khrennikov and Kozyref (2006).

\item Wavelet basis on $L^2(\Q_p)$ using translation operators defined 
on the compact, open subgroup $\Z_p$.  (It is interesting to note that
$\Z^m$ is discrete; and that the quotient $\R^m/\Z^m$ is compact.  In 
contrast to this, $\Z_p$ is compact; and the quotient $\Q_p/\Z_p$ is
discrete.)
\item Discussed is a wavelet basis on $L^2(G)$, for a group $G$, using
translation operators defined on a discrete subgroup, or discrete lattice.
\item Finally the central theme of Benedetto (2004) is a wavelet basis 
on $L^2(G)$ where $G$ is a locally compact abelian group, using translation 
operators defined on a compact open subgroup (or operators that can be used
as such on a compact open subgroup); and with definition of an 
expansive automorphism replacing the traditional use of dilation.  
\end{itemize}

A motivation for the work of Benedetto (2004) and Benedetto and Benedetto
(2004) is laying the groundwork for  the wavelet transform on 
the adelic numbers (see Appendix 3).   In this work we are content to 
be less ambitious in regard to number systems -- below we focus on a 
particular p-adic encoding of dendrograms; and we are also less ambitious in 
regard to wavelet functions -- staying resolutely with the Haar wavelet
in this work.  Our motivation is due to our application drivers.

Locally compact abelian groups (LCAG) 
are a way to take Fourier analysis (hence a particularly important class 
of harmonic analysis because so versatile) into more general settings than
e.g.\ the reals (although the reals also form a non-compact, but locally 
compact, abelian group).  

The duals of members of a locally compact 
abelian group,
defined as unitary multiplicative characters, $x \longrightarrow 
e^{-2 \pi i x}$, also form a locally compact abelian group (Knapp, 1996).  
The duality pairing $G$ and $\widehat{G}$ allows for an isometry between 
$L^2(G)$ and $L^2(\widehat{G})$ (Antoine et al., 2000).  

Fourier analysis is the study of real square integrable functions that 
are invariant under the group of integer translations (see Foote et al., 
2000a), while abstract harmonic analysis is the study of functions on 
more general topological groups that are invariant under a (closed) 
subgroup.  

It is interesting to compare some global properties of our approach
relative to the Fourier transform approach applied to decision trees in
Kargupta and Park (2004).
The Fourier transform lends itself well to a frequency
spectrum analysis of binary decision vectors, and the latter can be of
importance for supervised classification.  On the other hand, our work
makes use of binary trees but in the framework of unsupervised
classification.  The wavelet transform
shares with the Fourier transform the property
that frequency spectral information is
determined from the data; and the wavelet transform additionally determines
spatial or resolution scale information from the data.  We have found
the wavelet transform, as described in this article, to be appropriate
for the type of input data that we have considered.  In general terms,
both we in this work, and Kargupta and Park (2004),
have as objectives the filtering and compression of data.

We need affine group action for the wavelet transform, and we have seen
above (section \ref{discrete}, ``Wavelet transform on discrete fields'') 
that $\Z_p$ affords us this; but for an 
arbitrary discrete field, and for an arbitrary locally compact abelian 
group, it is tricky to find an affine group.   
Taking further the Flornes et al.\ (1994) 
work, Antoine et al.\ (2000) consider 
an infinite locally compact abelian group, $\mathcal{G}$; the 
restriction of $\mathcal{G}$ to a lattice $\Gamma \subset \mathcal{G}$; 
$\mathcal{A}$, an abelian semigroup; and the actions of $\mathcal{A}$ on 
$\ell^2(\Gamma)$.  Based on a pseudodilation (i.e., the product of a 
natural dilation by a convolution operator) the case of a continuous 
underlying signal is studied, i.e.\ the relation between the semigroup 
acting on $\ell^2(\Gamma)$, and a continuous affine group acting on 
$L^2(\R)$.  Spline functions  are again among the wavelet functions used 
(among which is the Haar wavelet function associated with the B-spline of 
order 1).  

The aspect of greatest interest to us here in the approach of 
Benedetto (2004), and Benedetto and Benedetto (2004), is to define 
wavelets on $L^2(G)$, with $G$ taken as the p-adic rationals, $\Q_p$, and
with the p-adic integers, $\Z_p$, on which we define translation-like 
operators.   Firstly, we are using therefore $L^2(\Q_p)$, i.e.\ 
functions defined on the rationals.  Secondly, we use a discrete group of
operators on $L^2(\Q_p)$ which are not in themselves translation operators,
but may be used in an analogous way.  The ``trick'' used is that the quotient
$\Q_p / \Z_p$ is discrete, and this will furnish the translation 
operators.  An expansive automorphism is also 
needed in this context, i.e.\ what we use in analogy with dilation.  

A number of alternatives for the subset of $\Q_p / \Z_p$ (more strictly 
the quotient of the group dual by the annhilator in the dual 
 of the compact open subgroup) are discussed by Benedetto (2000a).  Given
our application-driven interest, we will not pursue them further here.
What we will do, however, is to look at how the group-based approach of 
Benedetto (2000a), that for the most part assumes infinite sets, can be 
tailored for our algorithmic -- hence finite -- purposes.  

We will therefore look at how we can suitably encode any given dendrogram  
in terms of $\Q_p$ -- or indeed, as will be seen, in terms of $\Z_p$.  

Next we will move on to look at how a lattice-proxy is defined on our 
encoding, and thereby translation operators.

Finally, we will look at how an expansive automorphism can be replaced by 
expansive mapping in the finite and discrete case.  

In all of this, we follow the methodology described by Benedetto (2000a); 
but we restrict all aspects to the finite, discrete context.  

\subsection{The Wreath Product Group Corresponding to a Hierarchical Clustering} 

For the group actions, with respect to which we will seek 
invariance, we consider independent cyclic shifts of the 
subnodes of a given node (hence, at each level).  Equivalently
these actions are adjancency preserving permutations of 
subnodes of a given node (i.e., for given $q$, the permutations 
of $\{  q', q'' \}$.  Due to the binary tree, or strictly 
pairwise agglomerations represented by the hierarchy, the 
``adjacency'' property is trivial.  We have therefore cyclic 
group actions at each node, where the cyclic group is of order
2.  

The symmetries of $H$ are given by structured permutations of
the terminals.  The terminals will be denoted here by 
Term $H$. The full group of symmetries is summarized
by the following algorithm:

\begin{enumerate}
\item For level  $l = n - 1$ down to 1 do:
\item Selected node, $\nu \longleftarrow $ node at level $l$.
\item And permute subnodes of $\nu$.  
\end{enumerate}

Subnode $\nu$ is the root of subtree $H_\nu$.  We denote $H_{n-1}$ simply by
$H$.  For a subnode $\nu'$ undergoing a relocation action in step 3, the 
internal structure of subtree $H_{\nu'}$ is not altered.  

The algorithm described defines the automorphism group which is a 
wreath product of the symmetric group.  Denote the permutation at level 
$\nu$ by $P_\nu$.  Then the automorphism group is given by:
$$G = P_{n-1} \ \mathrm{wr} \ P_{n-2} \ \mathrm{wr} \ \dots \ \mathrm{wr} \ P_2
\  \mathrm{wr} \  P_1$$
where wr denotes the wreath product.  

Call Term $H_\nu$ the terminals that descend from the node at level $\nu$. 
So these are the terminals of the subtree $H_\nu$ with its root node at 
level $\nu$.  We can alternatively call Term $H_\nu$ the cluster associated
with level $\nu$.  

We will now look at shift invariance under the group action.  This amounts 
to the requirement for a constant function defined on Term $H_\nu, \forall 
\nu$.  A convenient way to do this is to define such a function on the set
Term $H_\nu$ via the root node alone, $\nu$.  By definition then we have a
constant function on the set Term $H_\nu$.  

Let us call $V_\nu$ a space of functions that are constant on Term $H_\nu$. 
Possible bases of $V_\nu$ that were considered in Paper I are:

\begin{enumerate}
\item Basis  vector with $| \mathrm{Term} H_\nu |$ components, with 0 values
except for value 1 for component $i$.   
\item Set (of cardinality $n = | \mathrm{Term} H_\nu |$) of $m$-dimensional
observation vectors.
\end{enumerate}

The constant function maps 
$$ L(\mathrm{Term} H) \longrightarrow V_\nu$$
where $L$ is the space of complex valued functions on the set Term $H$.  

Now we consider the resolution scheme arising from moving from \\
$\{ \mathrm{Term} H_{\nu'}, \mathrm{Term} H_{\nu''} \}$ to
$ \mathrm{Term} H_\nu $.  From the hierarchical clustering point of view it is 
clear what this represents: simply, an agglomeration of two clusters 
called Term $H_{\nu'}$ and Term $H_{\nu''}$, replacing them with a new 
cluster, Term $H_\nu$.  

Let the spaces of constant functions 
corresponding to the two cluster agglomerands be denoted $V_{\nu'}$ and 
$V_{\nu''}$.  These two clusters are disjoint initially, which motivates
us taking the two spaces as a couple: $(V_{\nu'}, V_{\nu''})$.  
In the same way, let the space of constant functions 
corresponding to node $\nu$ be denoted $V_\nu$.  

The multiresolution scheme uses a space of zero mean denoted $W_{\nu' \nu''}$
with mean defined on the couple of spaces, $(V_{\nu'}, V_{\nu''})$:

$$ V_\nu = (V_{\nu'}, V_{\nu''}) \oplus W_{\nu' \nu''} =
(V_{\nu'} \oplus W_{\nu' \nu''}, V_{\nu''} \oplus W_{\nu' \nu''}) $$

In considering spaces of constant functions, 
$V_{\nu'}$ and $V_{\nu''}$, we know that the support of these spaces are,
respectively, Term $H_\nu'$ and Term $H_{\nu''}$.  So if, instead of the 
space of zero mean denoted $W_{\nu' \nu''}$
where mean is defined on the couple of spaces, $(V_{\nu'}, V_{\nu''})$, we
considered the mean of the combined support, Term $H_\nu'  \cup$  
Term $H_{\nu''}$, then the result would be quite different.  We would, 
in fact, have a cluster-weighted mean value.  

\subsection{Example}

Let us exemplify a case that satisfies all that has been 
defined in the context of the wreath product invariance that we are
targeting.  It also exemplifies the algorithm discussed in depth in 
Paper I. Take the constant function on $V_{\nu'}$ to be $f_{\nu'}$.
Take the constant function on $V_{\nu''}$ to be $f_{\nu''}$. 
Then define the constant function on $V_{\nu}$ to be 
$(f_{\nu'} + f_{\nu''})/2$.  Next define the zero mean function on 
$W_{\nu' \nu''}$ to be:

$$w_{\nu'} = (f_{\nu'} + f_{\nu''})/2 - f_{\nu'}$$
in the support interval of $V_{\nu'}$, i.e.\ Term $H_{\nu'}$, and 
$$w_{\nu''} = (f_{\nu'} + f_{\nu''})/2 - f_{\nu''}$$
in the support interval of $V_{\nu''}$, i.e.\ Term $H_{\nu''}$.

Evidently $w_{\nu'} = - w_{\nu''}$.

\subsection{Inverse Transform}

Following on from 
the previous subsection, a demonstration that the algorithm allows 
for exact reconstruction of the data -- the inverse transform -- is as 
follows.  The constant function on Term $H$ corresponding to the root node
is $f_{n-1}$.  The two subnodes of the root node, at levels $\nu'$ and 
$\nu''$, are reconstructed from $f_{n-1} \pm w_{\nu'}$ (and, as we have seen,
we can either use $w_{\nu'}$ or $w_{\nu''}$).  We next look at the subtrees 
whose roots are given by the nodes (just considered) at levels $\nu'$ and
$\nu''$, and these subtrees are necessarily disjoint.  All subnodes of these
currently selected nodes are reconstructed using the same algorithm.  This 
procedure is iteratively continued until the terminals have been dealt with.

\subsection{Link with Agglomerative Hierarchical Clustering Algorithms}

Comparison with traditional clustering criteria is considered next.
It is clear why agglomerative
levels are very problematic if used for choosing a
good partition in a hierarchical clustering tree: 
they increase with agglomeration, simply because the cluster
centers are getting more and more spread out as the sequence of agglomerations
proceeds.  Directly using these agglomerative levels has been a way
to derive a partition for a very long time.  An early reference is Mojena
(1977).  To see how the wavelet transform 
used by us leads to a very different outcome, see Paper I, where we
describe use of the norms of the $w_{\nu'}$ vectors.   

  
When we consider agglomerative hierarchical clustering algorithms it is 
clear that (i) clusters are often defined in terms of center of gravity,
or mean; and (ii) this allows for defining a (vector) difference term between
a cluster and its immediate sub-clusters.  It is also clear that the 
Haar wavelet algorithm is close to the method known as median or Gower's or
WPGMC -- weighted pair group method using centroids: see table, p.\ 68, of
Murtagh (1985).  

We could also cater for other agglomerative criteria, subject to storing 
the cluster cardinality values, and develop an algorithm that is close to the 
Haar wavelet one.  Constant functions on spaces $V_{\nu'}$ and $V_{\nu''}$ 
remain just as before.  The zero mean functions on space $W_{\nu' \nu''}$ 
would now be generalized to weight mean (with weights given by cluster
cardinalities).  Viewed in this light, our work has led us to develop a new
storage structure for hierarchical clustering trees, which is particularly 
beneficial for data filtering objectives.

The novelty of our work resides in two areas: (i) we have shown the close
association between two classes of multiple resolution data analysis 
approaches, agglomerative hierarchical clustering algorithms and 
the wavelet transform; and (ii) our motivation is not at all to construct 
hierarchical clusterings in a new way but rather to illuminate further 
inherent ultrametric properties of data (cf.\ Murtagh, 2004).  

\section{An Algebraic Representation of a Hierarchy}

\subsection{Introduction}

The dendrogram wavelet transform has been seen to be a set of applications of 
a function 
applied to set members, or cluster members, associated with nodes of the 
hierarchy $H$.  
The non-singleton clusters comprise the set $Q = \{ q_1,  q_2, \dots , q_{n-1}
\}$.  The wavelet function is applied in turn to $q_1, q_2, \dots $.  In 
this paper, we are using the Haar 
wavelet function.  But we could well use others (e.g., Altaisky, 2004, 
uses the Morlet wavelet).

With each $q \in Q$ there is an associated level function, $\nu : q 
\longrightarrow \R^+$, which induces a total order on $Q$.  We will show that 
the application of the wavelet function to this sequence of clusters 
is ``dilatary'' or ``expansive'' in two different ways.  

We will return to what these two different ways are in a moment.  
The hierarchy $H_{\nu = 0}$ contains an inceasingly embedded 
sequence of subsets (corresponding to increasingly pruning the branches of
the tree; Bouchki, 1996).  We have: 
$H_{\nu = 0} \supset H_{\nu 1} \supset H_{\nu 2} \supset \dots 
\supset H_{\nu (n-1)} $.  Define Term as the set of terminal nodes of a 
hierarchy, and Card the cardinality of a set.  Then Term$(H_{\nu = 0}) = I$ 
with Card$(I) = n$.  Card(Term$(H_{\nu 1})$) $ = n-1 $. 
Card(Term$(H_{\nu 2})$) $ = n-2 $.  $\dots$  
Card(Term$(H_{\nu (n-1)})$) $ = 1 $.  Each application of the wavelet 
function is to the minimal (non-singleton) cluster (again see Bouchki, 1996)
in each $_{\nu k}$.  Below, we will see how we {\em promote} all 
$q \in H_{\nu k}$ to the corresponding $q \in H_{\nu (k+1)}$ by multiplying
clusters (in a particular algebraic representation) by $1/p$, which is
of norm $p$.   

The ``dilatory'' or ``expansive'' character of our sequence of operations,
viz., application of the wavelet function, comes from (i) the sequence of
embedded subsets of $H$ -- so we still operate on a cluster, but the data 
on which we work becomes smaller; or (ii) the sequence of levels 
at which we work is derived from repeatedly taking the product with $1/p$ of
norm $p$.  

In order to introduce this product with $1/p$ of norm $p$ we first 
describe the p-adic algebraic representation of the hierarchy.  In this 
representation, clusters including singletons, have an operator, 
denoted $\oplus$.  This operator allows clusters to be defined.  Next, 
we have a null element in $Q$ in this algebraic representation, and a 
norm of each $q \in Q$.  Hence for $q', q'' \in Q$, we can define 
$q' \oplus q''$.  We have: $\exists q \in Q$ s.t. $q = 0$.  Finally,
$\forall q \in Q$ we have $\| q \|$.  

\subsection{$H$ Expressed p-Adically: p-Adic Encoding of a Dendrogram}

We will introduce now the one-to-one mapping of clusters (including 
singletons) in $H$ into a set of p-adically expressed integers (a fortiori,
rationals, $\Q_p$).   
The field of p-adic numbers is the most important example of
ultrametric spaces.  
Addition and multiplication of p-adic integers, $\Z_p$, 
are well-defined.  Inverses
exist and no zero-divisors exist.  

A terminal-to-root traversal in a dendrogram or binary rooted tree
is defined as follows.  We use  
the path $x \subset q \subset q' \subset q'' \subset \dots q_{n-1}$, where
$x$ is a given object specifying a given terminal, and $q, q', q'', \dots$
are the embedded classes along this path, specifying nodes in the
dendrogram.  The root node is specified by the class $q_{n-1}$ 
comprising all objects.  

A terminal-to-root traversal is the shortest path between the given terminal
node and the root node, 
assuming we preclude repeated traversal (backtrack) 
of the same path between any two nodes.  

\begin{figure}
\centering
\includegraphics[width=9cm,angle=270]{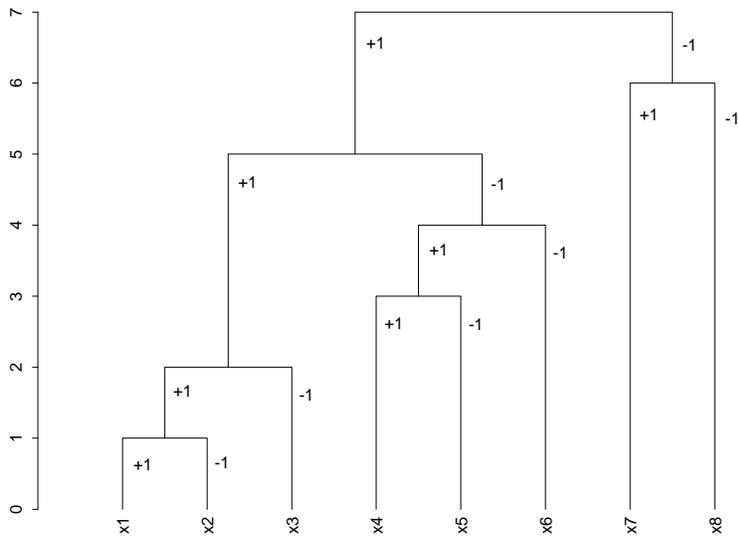}
\caption{Labeled, ranked dendrogram on 8 terminal nodes,
$x_1, x_2, \dots , x_8$.  Branches are labeled 
$+1$ and $-1$. Clusters are: $q_1 = \{ x_1, x_2 \}, 
q_2 = \{ x_1, x_2, x_3 \}, q_3 = \{ x_4, x_5 \}, q_4 =  \{ x_4, x_5, 
x_6 \}, q_5 = \{ x_1, x_2, x_3, x_4, x_5, x_6 \}, 
q_6 = \{ x_7, x_8 \}, q_7 = \{ x_1, x_2, \dots , x_7, x_8 \}$.}
\label{fig2}
\end{figure}

By means of terminal-to-root traversals, we define the following 
p-adic encoding of terminal nodes, and hence objects, in Figure \ref{fig2}.

\begin{eqnarray*}
x_1: & + 1 \cdot p^1 + 1 \cdot p^2 + 1 \cdot p^5 + 1 \cdot p^7  \\
x_2: & - 1 \cdot p^1 + 1 \cdot p^2 + 1 \cdot p^5 + 1 \cdot p^7  \\
x_3: & - 1 \cdot p^2 + 1 \cdot p^5 + 1 \cdot p^7  \\
x_4: & + 1 \cdot p^3 + 1 \cdot p^4 - 1 \cdot p^5 + 1 \cdot p^7   \\
x_5: & - 1 \cdot p^3 + 1 \cdot p^4 - 1 \cdot p^5 + 1 \cdot p^7   \\
x_6: & - 1 \cdot p^4 - 1 \cdot p^5 + 1 \cdot p^7  \\
x_7: & + 1 \cdot p^6 - 1 \cdot p^7   \\
x_8: & - 1 \cdot p^6 - 1 \cdot p^7 
\end{eqnarray*}

If we choose $p = 2$ the resulting decimal equivalents could be identical:
cf.\ contributions based on $+1 \cdot p^1$ and $-1 \cdot p^1 + 1 \cdot p^2$.
Given that the coefficients of the $p^j$ 
terms ($1 \leq j \leq 7$) are in the set $\{ -1, 0, +1 \}$
(implying for $x_1$ the additional terms: $+ 0 \cdot p^3 + 0 \cdot p^4
+ 0 \cdot p^6$), 
the coding based on $p = 3$ is required to avoid ambiguity among 
decimal equivalents. 

A few general remarks on this encoding follow.  For the labeled ranked
binary trees that we are considering, we require the labels $+1$ and $-1$
for the two branches at any node.  Of course we could interchange 
these labels, 
and have these $+1$ and $-1$ labels reversed at any node. 
By doing so we will have different p-adic codes for the objects, $x_i$.  

The following properties hold: (i) {\em Unique encoding:} the 
decimal codes for
each $x_i$ (lexicographically ordered) are unique for $p \geq 3$; 
and (ii) {\em Reversibility:} the dendrogram can be uniquely reconstructed
from any such set of unique codes.  

The p-adic encoding defined for any object set 
 above can be expressed as follows for any object $x$ 
associated with a terminal node:

\begin{equation}
x = \sum_{j=1}^{n-1} c_j p^j  \mbox{    where } c_j \in \{ -1, 0, +1 \}
\label{eqn1}
\end{equation}

In greater detail we have: 

\begin{equation}
x_i = \sum_{j=1}^{n-1} c_{ij} p^j  \mbox{    where } c_{ij} 
\in \{ -1, 0, +1 \}
\label{eqn1b}
\end{equation}

Here $j$ is the level or rank (root: $n-1$; terminal: 1), and $i$ is an 
object index.  

In our examples we have used: $a_j = +1$ for a left branch
(in the sense of Figure \ref{fig2}), $= -1$ for a right
branch, and $= 0$ when the node is not on the path from that particular 
terminal to the root.  

A matrix form of this encoding is as follows, where 
$\{ \cdot \}^t$ denotes the transpose of the vector.  

Let $\mathbf{x}$ be the column vector $\{ x_1 \ x_2 \ \dots x_n \}^t$.

Let $\mathbf{p}$ be the column vector $\{ p^1 \ p^2 \ \dots p^{n-1} \}^t$.

Define a characteristic matrix $C$ of the branching codes, $+1$ and $-1$, 
and an absent or non-existent branching given by $0$,
as a 
set of values $c_{ij}$ where $i \in I$, the indices of the object set; and
$j \in \{ 1, 2, \dots , n-1 \}$, the indices of the dendrogram levels 
or nodes ordered 
increasingly.  For Figure \ref{fig2} we therefore have: 

\begin{equation}
C = \{ c_{ij} \} = 
\left(
\begin{array}{rrrrrrr}
1  & 1 & 0 &  0 & 1 &  0 & 1 \\   
-1 & 1 & 0 &  0 & 1 & 0 & 1 \\
0  & -1 & 0 &  0 & 1 & 0 & 1 \\
0  &  0 & 1 &  1 & -1 & 0 & 1 \\
0  & 0  & -1 & 1 & -1 & 0 & 1 \\
0  & 0 & 0  & -1 & -1 & 0 & 1 \\
0  & 0 & 0  & 0  & 0 & 1 & -1 \\
0  & 0 & 0  & 0  & 0 & -1 & -1 
\end{array}
\right)
\label{eqn2}
\end{equation}

For given level $j$, $\forall i$, the absolute values $| c_{ij} |$ 
give the membership function either by node, $j$, which is therefore 
read off 
columnwise; or by object index, $i$ which is therefore read off rowwise.

The matrix form of the p-adic encoding is:

\begin{equation}
\mathbf{x} = C \mathbf{p} 
\label{eqn3}
\end{equation}

Here, {\bf x} is the decimal encoding, $C$ is the matrix with dendrogram
branching codes and {\bf p} is the vector of powers of a fixed integer
(usually, more restrictively, fixed prime) $p$.  

The tree encoding exemplified in Figure \ref{fig2}, and defined with 
 coefficients
in equations (\ref{eqn1}) or (\ref{eqn1b}), 
(\ref{eqn2}) or (\ref{eqn3}), with labels  $+ 1 $ and 
$ - 1$ is not commonly used:  
zero and one labels are more common.  We required 
the $\pm 1$ labels, however, to fully cater for the ranked nodes (i.e.\ the 
total order, as opposed to a partial order, on the nodes).  

We can 
consider the objects that we are dealing with to have equivalent integer 
values.  To show that, all we must do is work out decimal equivalents of 
the p-adic expressions used above for $x_1, x_2, \dots $.  As noted 
in Gouv\^ea (2003), we have equivalence between: a p-adic number; a 
p-adic expansion; and an element of $\Z_p$ (the p-adic integers).
The coefficients used to specify a p-adic number, Gouv\^ea (2003) notes 
(p.\ 69), ``must be taken in a set of representatives of the class modulo
$p$.  The numbers between 0 and $p-1$ are only the most obvious choice for
these representatives.  There are situations, however, where other choices
are expedient.''

\subsection{P-adic Dendrogram Addition and Multiplication}

As noted already the wavelet basis on $L^2(\R^m)$ is often 
induced from the discrete subgroup, $\Z^m$ . Now for a discrete subgroup 
we use the dendrogram, $H$.  The addition operation on the group 
$H$ will now be explored. 

In order to define a group structure on the p-adic encoded objects, 
we require an addition operation.  We do not ``carry and add'' in the 
traditional way because this does not make sense in this context.  
Instead we define the following ``average and threshold'' operation for 
any coefficients (of values of $\mathbf{p}$, as used in equations 
\ref{eqn1b} or \ref{eqn3}).  We define the following compositions
for such coefficients. 

\begin{equation}
\begin{array}{cccccr} 
 + & 1 & + & 1 & \longrightarrow & + 1 \\ 
 - & 1 & - & 1 & \longrightarrow & - 1 \\ 
 + & 1 & - & 1 & \longrightarrow & 0 \\ 
 - & 1 & + & 1 & \longrightarrow & 0 \\ 
 + & 1 & \pm & 0 & \longrightarrow & 0 \\ 
 - & 1 & \pm & 0 & \longrightarrow & 0
\end{array}
\label{eqn4}
\end{equation}

Examples from the encoding defined above for $x_1, x_2, \dots $ 
(again with reference to Figure \ref{fig2}, and equations 
\ref{eqn1} or \ref{eqn1b}, \ref{eqn2} or \ref{eqn3}) follows.

\medskip

$x_1 \oplus x_2 = + 1 \cdot p^2 + 1 \cdot p^5 + 1 \cdot p^7 $

$x_1 \oplus x_3 = + 1 \cdot p^5 + 1 \cdot p^7 $

$x_1 \oplus x_7 = 0$

$x_3 \oplus x_6 = + 1 \cdot p^7 $

$x_5 \oplus x_8 = 0 $

\medskip

Informally: in the tree, this addition operation only retains non-zero terms 
for nodes in the tree strictly {\em above} the first (i.e.\ lowest level) 
cluster within which the two objects find themselves.  This means that
if the two objects only find themselves together for the first time 
in the same cluster that contains all objects then the result of the 
addition operation is 0.  

Let us use our ``average and threshold'' operation, which we are
using as a customized addition, to define clusters.  We will do so
by example, taking Figure \ref{fig2} as our case study.  We will call the 
clusters, ranked by increasing node level, $q_1, q_2, \dots $ as used in the 
caption of Figure \ref{fig2}.

\medskip

$ q_1 = x_1 \oplus x_2 = + 1 \cdot p^2 + 1 \cdot p^5 + 1 \cdot p^7 $

$ q_2 = q_1 \oplus x_3 = + 1 \cdot p^5 + 1 \cdot p^7 $

$ q_3 = x_4 \oplus x_5 = + 1 \cdot p^4 - 1 \cdot p^5 + 1 \cdot p^7 $

$ q_4 = q_3 \oplus x_6 = - 1 \cdot p^5 + 1 \cdot p^7 $

$ q_5 = q_2 \oplus q_4 =  + 1 \cdot p^7 $

$ q_6 = x_7 \oplus x_8 = - 1 \cdot p^7 $

$ q_7 = 0 $

\medskip

The trivial cluster containing all $n$ objects, $q_{n-1}$, is of
value 0 in this representation.  

\bigskip

\noindent
Definition of Null Element:

\medskip

\noindent
On the dendrogram $H$, the set $q_{n-1} = I$ is the null
element when using our p-adic encoding (given in 
definitions (\ref{eqn1b}) and (\ref{eqn3})) 
and addition operation (\ref{eqn4}).  

\bigskip

Defining p-adic notation for clusters in this way allows us to define
norms of clusters; or to define p-adic distances between clusters; or 
indeed to define p-adic distances between clusters and objects (singletons,
terminals).  We will look at these in subsection \ref{sect54} below.

For completeness we will provide a definition of p-adic dendrogram
multiplication.
Take $ x = \sum_j c_j p^j$ and let $ y = \sum_j c'_j p^j$.  The product
operation is defined on the formal (Laurent) power series as: 

\begin{equation}
x y = \left(    \sum_j c_j p^j \right) \left( \sum_{j'} c'_{j'} p^{j'} 
\right)  =
\sum_{jj'} c_j c'_{j'} p^{j + j'}
\end{equation}
with restriction to the term in $p^{n-1}$.
P-adic dendrogram multiplication will be used below in the definition of
the expansive operator: this is multiplication by $1/p$.

\subsection{P-adic Distance and Norm on a Dendrogram}
\label{sect54}

Thus far, we have been concerned with an analytic framework.  Now
we will induce a metric topology on $H$.  

To find the p-adic distance, we look for the term $p^r$ in the p-adic 
codes of the two objects, where $r$ is the lowest level such that the 
absolute values of the coefficients of $p^r$ are equal.

Let us look at the set of p-adic codes for $x_1, x_2, \dots $ above
(Figure \ref{fig2}), to 
give some examples of this. 

\medskip

For $x_1$ and $x_2$, we find the term we are looking for to be $p^1$, and 
so $r = 1$.  

For $x_1$ and $x_5$, we find the term we are looking for to be $p^5$, and 
so $r = 5$.

For $x_5$ and $x_8$, we find the term we are looking for to be $p^7$, and 
so $r = 7$.

\medskip

Having found the value $r$, the distance is defined as $p^{-r}$. 

See, inter alia, Benz\'ecri (1979), and Gouv\^ea (2003), for this 
definition of ultrametric distance.

Examples based on Figure \ref{fig2}: 

\medskip

$| x_1 - x_2|_p = | x_2 - x_1 |_p = p^{-1} $ since $r = 1$.     

$| x_1 - x_4|_p = | x_4 - x_1 |_p = p^{-5} $ since $r = 5$.  

$| x_3 - x_6|_p = | x_6 - x_3 |_p = p^{-5} $ since $r = 5$.

\medskip

Examples for clusters from Figure \ref{fig2}: 

\medskip

$| q_1 - q_3|_p = |q_3 - q_1|_p = p^{-5}$.  

$| q_2 - q_6|_p = |q_6 - q_2|_p = p^{-7}$.

\medskip

We take for a singleton 
object $r = 0$, and so the norm of an object is always 1.
We therefore define the p-adic norm, $| . |_p$, of an object corresponding 
to a terminal node in the following way: for any object, $x$, $| x |_p = 1$.  


The norm of a non-singleton cluster is defined analogously.  
It is seen to be strictly smaller.  
We have: $ | q_2 |_p = p^{-2}; | q_4 |_p = p^{-4}$.  

For the expansive operator that we use for  dilation, 
we will consider
product with $1/p$.  The norm associated with this operator is seen to be 
$| 1/p |_p = | p^{-1} |_p = p^{-(-1)} = p$.  

The operator given by multiplication by $1/p$ therefore has norm 
or modulus $p$.  

The p-adic norm, or p-adic valuation, satisfies the following
properties (Schikhof, 1984): 

\begin{enumerate}
\item $| x |_p \geq 0; | x |_p = 0 $ iff $ x = 0 $
\item $| x + y |_p \leq \mbox{max} ( | x |_p, | y |_p )$
\item $| xy |_p = | x |_p | y |_p $
\end{enumerate}

We also have: $| q |_p \leq 1$ with equality only if $q$ is a singleton.

\subsection{Modified Dilation Operation: Multiplication by $1/p$}





Consider the set $\{ x_i | i \in I \}$ with its p-adic coding considered 
above.  Take $ p = 2$.  (Non-uniqueness of corresponding decimal codes is 
not of concern to us now, and taking this value for $p$ is without any 
loss of generality.)
Multiplication of 
$x_1 = + 1 \cdot 2^1 + 1 \cdot 2^2 + 1 \cdot 2^5 + 1 \cdot 2^7 $ by 
$1/p = 1/2$ gives: $  + 1 \cdot 2^1 + 1 \cdot 2^4 + 1 \cdot 2^6$.  Each
level has decreased by one, and the lowest level has been lost.  
Subject to the lowest level of the tree being lost, the form of the tree
remains the same.  By carrying
out the multiplication-by-$1/p$ operation on all objects, it is seen that 
the effect is to rise in the hierarchy by one level.   

Let us call product with $1/p$ the operator $A$.  The effect of losing the 
bottom level of the dendrogram means that either (i) each cluster (possibly
singleton) remains the same; or (ii) two clusters are merged.  Therefore 
the application of $A$ to all $q$ implies a subset relationship between 
the set of clusters $\{ q \}$ and the result of applying $A$, $\{ A q \}$.  


Repeated application of the operator $A$ gives $A q$, $A^2 q$,
$A^3 q$, $\dots$.  Starting with any singleton, $i \in I$, this gives 
a path from the terminal to the root node in the tree.  Each such
path ends with the null element, as a result of the Null Element definition
(section 3).  Therefore the intersection of the paths equals the 
null element.   

Benedetto and Benedetto (2004) discuss $A$ as an expansive
automorphism of $I$, i.e.\ form-preserving, and locally expansive.   

Some implications of Benederro and Benedetto's (2004) 
expansive automorphism follow.

For any $q$, let us 
take $q, A q, A^2 q, \dots$ as a sequence of open subgroups of 
$I$, with $q \subset A q \subset A^2 q \subset \dots$, and $I = 
\bigcup \{ q, A q, A^2 q, \dots \} $.  This is termed an inductive sequence 
of $I$, and $I$ itself is the inductive limit (Reiter and Stegeman, 2000).

Each path defined by application of the expansive automorphism 
defines a spherically complete system (Schikhof, 1984; Gaji\'c, 2001),
which is a 
formalization of well-defined subset embeddedness.  

We now return to our starting point, the Haar algorithm given in section 
1.4.  We apply the averaging and differencing operations to each cluster
in sequence.  But now, after doing this for cluster $q$, we apply the 
operator $A$, i.e.\ the $1/p$ product to the p-adic representation 
of the dendrogram.  This causes us to move up a level.  This is our 
enhanced concept of dilation, which we apply to the dendrogram, where we
keep the same averaging and differencing operations applied to the 
 cluster in sequence.  

\section{Wavelet Bases from the Wreath Product Group}
\label{dengroup}

In our case we are looking for a new basis for $L^2(G)$ where $G$ is the 
set of all equivalent representations of a hierarchy, $H$, on $n$ terminals.
Denoting the level index of $H$ as $\nu$ (so $\nu : H \longrightarrow \R^+$, 
where $\R^+$ are the positive reals), and $\nu = 0$ is the level index 
corresponding to the fine partition of singletons, then this hierarchy will
also be denoted as $H_{\nu = 0}$.  Let $I$ be the set of observations.  
Let the succession of clusters associated with nodes in $H$ be denoted
$Q = \{ q_1, q_2, \dots , q_{n-1} \}$.   
We have $n-1$ non-singleton nodes in $H$, associated with the clusters, $q$. 
At each node we can interchange left and right subnodes.  Hence we have 
$2^{n-1}$ equivalent representations of $H$, or, again, members in the 
group, $G$, that  we are considering.  

So we have the group of equivalent dendrogram representations on 
$H_{\nu = 0}$.  We have a series of subgroups, $H_{{\nu}_k} \supset 
H_{{\nu}_{(k+1)}}$, for $0 \leq k <  n-1$.
Symmetries are given by permutations at each level, $\nu$, of hierarchy $H$.
Collecting these furnishes a group of symmetries on the terminal set of any 
given (non-terminal) node in $H$. 
 
The practical application arises through identifying the $n$ terminal 
nodes with (i) $m$-dimensional vectors, or (ii) $n$-dimensional hypercube
vertices.  On the latter sets of vectors we can also consider an associated
permutation representation.  

Parenthetically, we note that the permutation representation is known 
as the alternating or zig-zag permutations and are counted by the Andr\'e
or Euler numbers (Murtagh, 1984a; sequence A000111 in Sloane, 2005).  

In this work we ignore another form of 
equivalent representation, i.e.\ that arising from two or more level values
being identical: $\nu_k = \nu_{k+1}$ for some $0 \leq k \leq n-1$.  This means
that successive nodes can be interchanged.  This situation happens when we
have equilateral triangles in the ultrametric space, as opposed to triangles
that are strictly isosceles with small base.  

At each non-singleton cluster, $q$, we define a (trivial) affine 
group on $(q', q'')$.  The group is defined on $Q = \{ q_\nu 
| \nu = 1, 2, \dots n-1 \}$.

Foote et al.\ (2000a) consider group actions on spherically homogeneous 
rooted trees.  The use of the latter is as a quadtree in 2D image 
processing.  (An image is recursively decomposed into spatially 
homogeneous quadrant covering regions; and this decomposition is represented
as a quadtree.  For 3D image volumes, the data structure becomes an 
octree.)  Just like for us, the quadtree nodes can ``twiddle'' around their
offspring nodes but, because of the image regions, group action amounts to
cyclic shifts or adjacency-preserving permutations of the offspring nodes.
The relevant group in this case is referred to as the wreath product group.

\section{Matrix Interpretation of the  Haar Dendgrogram Wavelet Transform}

\subsection{The Forward Transform}

Consider any hierarchical clustering, $H$, represented as a binary rooted
tree. For each cluster
$q''$ with offspring nodes $q$ and $q'$, we define $s(q'')$ through 
application of the low-pass filter 
$\left(
\begin{array}{r}
 \frac{1}{2} \\  
\frac{1}{2} 
\end{array}
\right)
$:

\begin{equation}
s(q'') = \frac{1}{2} \left( s(q) + s(q') \right) = 
\left(
\begin{array}{r}
0.5 \\
0.5
\end{array}
\right)^t
\left(
\begin{array}{c}
s(q) \\
s(q')
\end{array}
\right)
\end{equation}

The application of the low-pass filter is carried out in order of 
increasing node number (i.e., from the smallest non-terminal node, 
through to the root node).  

Next for each cluster
$q''$ with offspring nodes $q$ and $q'$, we define detail coefficients 
$d(q'')$ through 
application of the band-pass filter $\left( 
\begin{array}{r}
\frac{1}{2} \\ 
-\frac{1}{2} 
\end{array}
\right)$: 

\begin{equation}
d(q'') = \frac{1}{2} ( s(q) - s(q') ) = 
\left(
\begin{array}{r}
0.5 \\
-0.5
\end{array}
\right)^t
\left(
\begin{array}{c}
s(q) \\
s(q')
\end{array}
\right)
\end{equation}

Again, increasing order of node number is used for application of this
filter.  See Paper I for further details. 

\subsection{The Ultrametric Case}

We now return to the issue of how we start this scheme, i.e.\ how we 
define $s(i)$, or the ``smooth'' of a terminal node.  We have distinguished
above in section \ref{sect2} between:

\begin{enumerate}
\item $H$ as representing an ultrametric set of relations, 
\item $H$ as representing an embedded set of sets.
\end{enumerate}

For case 1 we take $s(i)$ as the $m$-dimensional observation vector 
corresponding to $i$.  So, taking all $n$ vectors $s(i)$ we have the 
initial data matrix $X$ of dimensions $n \times m$.  

Then for our set of $n$ points in $\R^m$ given in the form of 
matrix $X$ we have:

\begin{equation}
X = C D + S_{n-1}
\end{equation}

\noindent
where $D$ is the matrix collecting all
 wavelet projections or detail coefficients, $d$.  
The dimensions of $C$ are: $n \times (n-1)$ (see definition 
(\ref{eqn2})).  The dimensions of $D$ are 
$(n-1) \times m$. 

If $s_{n-1}$ is the final data smooth, in the limit for very large $n$ 
a constant-valued $m$-component vector, then let $S_{n-1}$ be 
the $n \times m$
matrix with $s_{n-1}$ repeated on each of the $n$ rows.  

Consider the $j$th coordinate of the $m$-dimensional observation 
vector corresponding to $i$.  
For any $d(q_j)$ we have: $ \sum_k d(q_j)_k = 0 $, i.e.\ the detail 
coefficient vectors are each of zero mean.


To recapitulate we have: 

$X$ is of dimensions $n \times m$.

$C$ is of dimensions $n \times (n-1) $.

$D$ is of dimensions $(n-1) \times m $.

$S_{n-1}$ is of dimensions $ n \times m$.

\subsection{The Case of Embedded Set of Sets}

We have distinguished between 

\begin{enumerate}
\item $H$ as representing an ultrametric set of relations,
\item $H$ as representing an embedded set of sets.
\end{enumerate}

We now turn attention to the latter.

In this case we take $s(i)$ as an $n$-dimensional indicator vector 
corresponding to $i$.  So, taking all $n$ vectors $s(i)$ we have the 
initial data matrix $X$ which is none other than the $n \times n$ 
dimensional identity matrix.  We will write $X_{\mbox{ind}}$ for this 
identity matrix.  

The wavelet transform in this case is: $X_{\mbox{ind}} = C D + S_{n-1}$.

$X_{\mbox{ind}}$ is of dimensions $ n \times n$.

$C$, exactly as in case 1 (ultrametric case) is of dimensions 
$ n \times (n-1)$.

$D$, of necessity different in values from case 1, is of dimensions
$ (n-1) \times n$.

$S_{n-1}$, of necessity different in values from case 1, is of dimensions
$ n \times n$.

\subsection{The Inverse Transform}

In both cases considered (viz., ultrametric, and set of sets) the 
forward and inverse transforms are performed in the same way.  The
algorithms are identical -- the inputs alone differ.

The inverse transform allows exact reconstruction of the input data.
We begin with $s_{n-1}$.  If this root node has subnodes $q$ and $q'$, we
use $d(q)$ and $d(q')$ to form $s(q)$ and $s(q')$.


\subsection{Wavelet Filtering}

Setting wavelet coefficients to zero and then reconstructing the data 
is referred to as hard thresholding (in wavelet space) and this is also 
termed wavelet smoothing or regression.  See the companion paper, 
Paper I, for discussion and examples. 


\subsection{Hierarchic Wavelet Transform in Matrix Form}

We will look at the ultrametric case.  The matrix generalization of 
equation (\ref{eqn3}) is: 

\begin{equation}
X = C P 
\label{e612}
\end{equation}

Matrix $P$ is formed from the vectors $
\mathbf{p}$ of equation (\ref{eqn3}) by replicating rows.

Now the wavelet transform gives us: $X = C D + S_{n-1}$.  Each 
(replicated) row of
matrix 
$S_{n-1}$ is a particular measure of central tendency. 

Centering $X$ relative to this gives:  

\begin{equation}
X - S_{n-1} = C D
\label{e613}
\end{equation}

We conclude from the formal similarity of expressions 
(\ref{e612}) and (\ref{e613}): 
the initial p-adic encoding of our data vectors has been
mapped into a {\em wavelet encoding} by the wavelet transform. 

With reference to section \ref{dengroup}, we note that relation 
\ref{e613} furnishes a unique matrix representation of a dendrogram.

\section{Discussion and Conclusions}

Generalization to regular 
$p$-way trees, for $p > 2$, may also be considered. 
For $p = 3$ a natural wavelet function is derived from the triangle scaling
(Starck et al., 1998)
function, which is itself a convolution of a box function (the scaling
function defining the Haar transform, used in this article) with itself. 
The Haar scaling function used above was $(\frac{1}{2}, \frac{1}{2})$.  
Convolving this with 
itself gives then the scaling function $(\frac{1}{4}, \frac{1}{2}, 
\frac{1}{4})$.  Convolving the 
box function again with the triangle function gives the B$_3$ spline 
scaling function, $(\frac{1}{16}, \frac{1}{4}, \frac{3}{8}, 
\frac{1}{4}, \frac{1}{16})$, which is particularly 
natural for the analysis of a 5-way, $p = 5$, tree.  


A remark on implementation
follows: the 3-way tree is unfolded at each node into two 2-way trees.  
More generally any regular $p$-way tree is unfolded at each node into 
$p-1$ two-way branchings.  The wavelet transform algorithm described 
previously is then directly applied. 


We now look at other related work.

In Khrennikov and Kozyrev (2004) and Kozyrev (2001) 
the Haar wavelet transform, defined on binary 
trees, was also introduced and discussed.  Compared to the notation used
here, the descriptions are related though a p-adic change of variable
(viz., $\sum_0^\infty a_i p^i $ is mapped onto $\sum_0^\infty a_i 
p^{-i-1}$.)  

For the $p = 2$ case, a convenient notational expression is given by the 
Vladimirov operator (see Avetisov, Bikulov, Kozyrev and Osipov, 2002)
which is a modified differentiation 
operator.  The Vladimirov operator is a p-adically expressed derivative for 
an ultrametric space with {\em linearly} related hierarchical levels, $\nu$.
In Kozyrev (2001, 2003) it is shown how the eigenvalues of the 
Vladimirov operator are the Haar wavelets.  As a consequence, the hierarchical
Haar wavelet transform is a spectral analysis of the Vladimorov operator.

Our work differs from the works cited in the following way.  Firstly, these
other works deal with regular $p$-way trees.  Degeneracies are allowed,
which can cater for the irregular $p$-way trees that we have considered.
We have preferred to directly address the dendrogram data structure, given 
that it models observed data well.  
Secondly, these other works cater for infinite trees.  We have restricted
ourselves to a more curtailed problem, with the aim of having a 
straightforward implementation, and with the aim of targeting the analysis 
of practical, constructive data analysis problems.  

We have also been more focused in this work compared to the general 
setting described in comprehensive depth by Benedetto and Benedetto (2004). 

An important reason for considering dendrograms rather than infinite 
regular trees is that the former setting gives rise to (low order) 
polynomially bound algorithms for all operations; whereas the latter, 
in the general case, are not polynomially bound.  

A final path for future work will be noted.  The Haar wavelet transform 
on a dendrogram ($H$) gives us information on the rate of change of the 
clusters ($q$), with respect to the level index of each cluster ($\nu$). 
In a sense this Haar wavelet transform is the derivative of $H$ with respect
to $\nu$.  This perspective may be of benefit when dealing with the 
dynamics of ultrametric spaces (Avetisov et al., 2002, and references 
therein; Kuhlmann, 2002).


\section*{Appendix 1. Haar Wavelet Transform Used in Image/Signal Processing}

Classically, the Haar wavelet function basis for analysis of $L^2(\R^m)$ 
is determined by inducing 
the basis from an $m$-dimensional pixel (time step, voxel, etc.) grid,
$\Z^m$. Basis functions of a space denoted by $V_j$ are defined from a
scaling function $\phi$ as follows (Starck, Murtagh and Bijaoui, 1998):
 
\begin{equation}
\phi_{j,i}(x) = \phi(2^{-j} x - i) \ \ \ \  i = 0, \dots , 2^j -1 \ \ \ \ \ \ \
\mbox{with} \  \phi(x) = \left\{ 
\begin{array}{ll}
1 & \mbox{for $0 \leq x < 1$} \\
0 & \mbox{otherwise} 
\end{array}
 \right.
\end{equation}
 
The functions $\phi$ are all box functions, defined
on the interval $[0,1)$ and are piecewise constant on $2^j$ subintervals.
We can 
approximate any function in spaces $V_j$ associated with basis functions 
$\phi_j$, in a very fine manner for $V_0$ (in this case of 
$V_0$, all values), 
more crudely for $V_{j+1}$ and so on.  We consider the nesting of spaces, 
$\dots V_{j+1} \subset V_{j} \subset V_{j-1} \dots \subset V_0 $.  
Equation (1) directly leads to a dyadic analysis. 
 
Next we consider the orthogonal complement of $V_{j+1}$ in 
$V_{j}$, and call it
$W_{j+1}$.  The basis functions for $W_j$ are derived from the 
Haar wavelet.  We find
 
\begin{equation}
\psi_{j,i}(x) = \psi(2^{-j} x - i) \ \ \ \  i = 0, \dots , 2^j -1 \ \
\mbox{with} \  \psi(x) = \left\{ 
\begin{array}{rl}
1 & \mbox{$0 \leq x < \frac{1}{2} $} \\
-1 & \frac{1}{2} \leq x < 1 \\
0 & \mbox{otherwise} 
\end{array}
\right.
\end{equation}
 
This leads to the basis for $V_{j}$ as being equal to: 
the basis for $V_{j+1}$ 
together with the basis for $W_{j+1}$.  
In practice we use this finding like this:
we write a given function in terms of basis functions in $V_{j}$; then we
rewrite in terms of basis functions in $V_{j+1}$ and $W_{j+1}$; 
and then we rewrite
the former to yield, overall, an expression in terms of basis functions 
in $V_{j+2}$, $W_{j+2}$ and $W_{j+1}$.  
The wavelet parts provide the detail part, 
and the space $V_{j+2}$ provides the smooth part.  
 
For the definitions of scaling function and wavelet function in the case of 
the Haar wavelet transform, proceeding from the given signal, the spaces
$V_j$ are formed by averaging of pairs of adjacent values, 
and the spaces $W_j$ are formed by 
differencing of pairs of adjacent values.  
Proceeding in this direction, from the given signal,
we see that application of the scaling or wavelet functions involves 
downsampling of the data. The low-pass
filter is a moving average.  The high-pass filter is a moving difference.
Other low-
and high-pass filters are alternatively used to yield other 
wavelet transforms.

\section*{Appendix 2. Hierarchy, Binary Tree and Ultrametric Topology}

A hierarchy, $H$, 
is defined as a binary, rooted, unlabeled, node-ranked tree, also 
termed a dendrogram (Benz\'ecri, 1979; Johnson, 1967; Lerman, 1981; Murtagh,
1985). 
A hierarchy defines a set of embedded subsets of a given set, $I$.  However
these subsets are totally ordered by an index function $\nu$, which is a
stronger condition than the partial order required by the subset relation.
A bijection exists between a hierarchy and an ultrametric space.  

Let us show these equivalences between embedded subsets, hierarchy, and 
binary tree, through the constructive approach of inducing $H$ on a set
$I$. 

Hierarchical agglomeration on $n$ observation vectors, $i \in I$, involves 
a series of $1, 2, \dots , n-1$ pairwise agglomerations of 
observations or clusters, with the following properties.  A hierarchy 
$H = \{ q | q \in 2^I \} $ such that (i) $I \in H$, (ii) $i \in H \ \forall 
i$, and (iii) for each $q \in H, q^\prime \in H: q \cap q^\prime \neq 
\emptyset \Longrightarrow q \subset  q^\prime \mbox{ or }  q^\prime
 \subset q$.  Here we have denoted the power set of set $I$ by $2^I$.
An indexed hierarchy is the pair $(H, \nu)$ where the positive
function defined on $H$, i.e., $\nu : H \rightarrow \R^+$, satisfies: 
$\nu(i) = 0$ if $i \in H$ is a singleton; and (ii)  $q \subset  q^\prime
\Longrightarrow \nu(q) < \nu(q^\prime)$.  Here we have denoted the 
positive reals, including 0, by $\R^+$.
Function $\nu$ is the agglomeration
level.  Take  $q \subset  q^\prime$, let $q \subset q''$  
 and $q^\prime \subset q''$, and let $q''$ be the lowest level cluster for 
which this is true. Then if we define $D(q, q^\prime) = \nu(q'')$, $D$ is 
an ultrametric.  In practice, we start with a Euclidean or other 
dissimilarity, use some criterion such as minimizing the change in variance
resulting from the agglomerations, and then define $\nu(q)$ as the 
dissimilarity associated with the  agglomeration carried out.

\section*{Appendix 3: P-Adic Numbers}

P-adic numbers were introduced by Kurt Hensel in 1898.  The ultrametric
topology was introduced by Marc Krasner (1944), the ultrametric inequality
having been formulated by Hausdorff in 1934. 
 Essential motivation for the study of this area is
provided by Schikhof (1984) as follows.  Real and complex fields gave rise
to the idea of studying any field $K$ with a complete valuation $| . |$
comparable to the absolute value function.  Such fields satisfy the
``strong triangle inequality'' $| x + y | \leq \mbox{max} ( | x |,
| y | )$.  Given a valued field, defining a totally ordered Abelian group,
an ultrametric space is induced through $| x - y | = d(x, y)$.
Various terms are used interchangeably for analysis in and
over such fields such as p-adic, ultrametric, non-Archimedean, and isosceles.
The natural geometric ordering of metric valuations is on the real line,
whereas in the ultrametric case the natural ordering is a hierarchical
tree.  P-adic numbers,
which provide an analytic version of ultrametric topologies,
have a crucially important property resulting from Ostrowski's theorem:
Each non-trivial valuation on the field of the rational numbers is equivalent
either to the absolute value function or to some p-adic valuation
(Schikhof, 1984,
p.\ 22).  Essentially this theorem states that the rationals can be
expressed in terms of (continuous) reals, or (discrete) p-adic numbers, and
no other alternative system.

The p-adic numbers are base p numbers, where p is a prime number.  It can be
shown that the reals can be expressed as p-adic numbers where p is infinity.
The question then arises as to whether any one of p = 2, 3, 5, 7, 11, $\dots
, \infty$ can be preferred.  For want of justification to limit attention 
to one or a few values of p, taking them all gives rise to the adelic 
number system (Brekke and Freund, 1993).

\section*{Appendix 4: Some Properties of Ultrametric Spaces}

See elsewhere for the basic ultrametric inequality, and the triangle 
propert -- isosceles with small base or equilateral.  The following is 
based on Lerman (1981), chapter 0, part IV.  

Theorem 1: 
Every point of a circle in an ultrametric space is a center of the circle.

Proof 1: it suffices to consider the triangle $a, b, x$, where $a$ is the 
center of the given circle, $b$ is an element of this circle, and $x$ is an
element of the circle with the same radius but with center $b$.  This triangle
is isosceles.  From the triangle property the result follows.  

Corollary 1: Two circles of the same radius, that are not disjoint, are
overlapping.

Definition 1: A divisor of the ultrametric space, $E$, 
is an equivalence relation
$D$ satisfying $ \forall a, b, x, y, \in E: \ \ aDb \ \ \mbox{and}
\ \  (d(x,y) \leq d(a,b)) \Longleftrightarrow xDy$.  

Corollary 2: Circles of the same radius form a partition of the 
ultrametric set.  The corresponding equivalence is a divisor of the space.

Definition 2: A valuation of a divisor $D$ of the space $E$ is the 
number $\nu(D) = \mbox{sup}_{xDy} d(x,y)$.  

Corollary 3: If $D$ and $D'$ are two divisors in $E$, a finite metric space,
verifying $D \leq D'$, then $\nu(D) \leq \nu(D')$ and reciprocally.  

Theorem 2: If $C$ adn $C'$ are disjoint circles in $E$, the distance
$d(x,y)$ of an $x \in C$ and of an $y \in C'$ depends on $C$ and $C'$ only,
and not on $x$ and $y$.

Proof 2: Consider the triangles $x, \bar{x}, y$, where $\bar{x} \in C'$ and
apply the ultrametric triangle relationship.

Corollary 4: The quotient $E/D$ of an ultrametric space by a divisor 
is an ultrametric space.  The distance between two of its points ist 
strictly greater than $\nu(D)$ in the finite case.

Definition  3: An ultrametric proximity is a positive (possibly infinite) 
function $p: E \times E \longrightarrow \R++ \cup \{ +\infty \}$, verifying
(i) $p(y,x) = p(x,y)$, (ii) $p(x,y) = + \infty$ iff $x = y$; and (iii)
$p(x,z) \geq \mbox{min} ( p(x,y), p(y,z) )$.  

Corollary 5: If $d$ is an ultrametric distance, then $- \log d$ is an
ultrametric proximity.  If $p$ is an ultrametric proximity, then $\exp(-p)$
is an ultrametric distance.  

Theorem 3:  For an $n \times n$  matrix of positive reals, symmetric with 
respect to the principal diagonal, to be a matrix of distances associated
with an ultrametric distance on $E$, a sufficient and necessary condition
is that a permutation of rows and columns satisfies the following form 
of the matrix: 

\begin{enumerate}
\item Above the diagonal term, equal to 0, the elements of the same row 
are non-decreasing.
\item For every index $k$, if 
$$d(k, k+1) = d(k, k+2) =  \dots = d(k, k+ \ell + 1)$$
then 
$$d(k+1, j ) \leq d(k,j)  \mbox{  for  } k + 1 < j \leq k + \ell + 1$$
and 
$$d(k+1, j) = d(k, j) \mbox{  for  } j > k + \ell + 1$$
Under these circumstances, $\ell \geq 0$ is the length of the section 
beginning, beyond the principal diagonal, the interval of columns of 
equal terms in row $k$.  
\end{enumerate}

Proof 4: Follows from ultrametric triangle inequality.  See Lerman 
(1981), p. 50.  

Theorem 5: In an ultrametric topology, every ball is both open and closed 
(termed clopen).  

(The empty set and the universal set are both clopen.  The complement of a
clopen set is clopen.  Finite unions and intersections of clopen sets are
clopen.)  

From Chakraborty (2004): 

A {\em basic neighborhood} of $x$, of radius $r$, is the set 
$N(x,r) = \{ y \in X: d(x,y) < r \}$.  
An {\em open set}, $U \subset X$, is a union of basic neighborhoods,
i.e.\ $\forall x \in U, \exists r = r(x) > 0 $ s.t. $N(x,y) \subset
U$.  

Many sets are open and closed at the same time.  This property is 
relative to subspaces.  Let $(X,d)$ be a metric space and $Y \subset
X$.  If $y \in Y$ and $r > 0$, let $N_Y(y,r)$ denote the basic 
neighborhood of $y$ in $Y$, and $N_X(y,r)$ the basic neighborhood
of $y$ in $X$.  Then $N_Y(y,r) = N_X(y,r) \cap Y$.  It follow that 
a set $U \subset Y$ is open in $Y$ iff $\exists$ an open set $V$ in 
$X$ s.t. $U = V \cup Y$.  An analogous statement holds for closed 
sets.  If $U \subset Y \subset X$ then $U$ can be open (or closed) 
in $Y$ without being open (or closed) in $X$.

\section*{Appendix 5: Ultrametric Spaces are 0-Dimensional}

Informally, a set of points is of necessity 0-dimensional.  

From Chakraborty (2004): 

A {\em base} $B$ for the topology $T$ is such that $B \subset T$, 
and every element of $T$ is a union of elements from $B$.

A metric space $(X,d)$ is called 0-dimensional if $\forall x \in X, r
> 0, \exists U$, a set, which is clopen, and $x \in U \subset N(x,r)$.

Van Rooij (1978): a topology is 0-dimensional if it has a base consisting
of clopen sets.  I.e., if for every $ a \in X$ and for every closed 
$A \subset X$ that does not contain $a$, there exists a clopen set 
$U$ such that $a \in U, A \subset X \backslash U$.

\end{document}